\documentclass[english,aps,manuscript]{revtex4}
\usepackage[T1]{fontenc}
\usepackage[latin9]{inputenc}
\usepackage{amsmath}
\usepackage{amssymb}
\usepackage{graphicx}
\usepackage{esint}

\makeatletter
\@ifundefined{textcolor}{}
{%
 \definecolor{BLACK}{gray}{0}
 \definecolor{WHITE}{gray}{1}
 \definecolor{RED}{rgb}{1,0,0}
 \definecolor{GREEN}{rgb}{0,1,0}
 \definecolor{BLUE}{rgb}{0,0,1}
 \definecolor{CYAN}{cmyk}{1,0,0,0}
 \definecolor{MAGENTA}{cmyk}{0,1,0,0}
 \definecolor{YELLOW}{cmyk}{0,0,1,0}
}

\usepackage{babel}

\usepackage{babel}

\usepackage{babel}

\usepackage{babel}

\usepackage{babel}

\makeatother

\usepackage{babel}
\begin{document}

\title{Optimal control of light propagation and exciton transfer in arrays of molecular-like
noble-metal clusters}

\author{Polina G. Lisinetskaya}

\affiliation{Institut für Physikalische und Theoretische Chemie, Universität Würzburg,
D-97074 Würzburg, Germany}

\author{Roland Mitri\'{c}}

\email{roland.mitric@uni-wuerzburg.de}

\affiliation{Institut für Physikalische und Theoretische Chemie, Universität Würzburg,
D-97074 Würzburg, Germany}
\begin{abstract}
We demonstrate theoretically the possibility of optimal control of
light propagation in arrays constructed of sub-nanometer sized noble
metal clusters by using phase-shaped laser pulses and analyze the
mechanism underlying this process. The theoretical approach for simulation
of light propagation in the arrays is based on the numerical solution
of the coupled time-dependent Schrödinger equation and the classical
electric field propagation in an iterative self-consistent manner.
The electronic eigenstates of individual clusters and the dipole couplings
are obtained from \textit{ab initio} TDDFT calculations. The total
electric field is propagated along the array by coupling an external
excitation electric field with the electric fields produced by all
clusters. A genetic algorithm is used to determine optimal pulse shapes
which drive the excitation in a desired direction. The described theoretical
approach is applied to control of the light propagation in a T-shaped
structure built of seven Ag$_{8}$ clusters. We demonstrate that a
selective switching of light localization is possible in $\sim$ 5
nm sized cluster arrays which might serve as building block for novel
plasmonic devices with ultrafast operation regime. 
\end{abstract}
\maketitle

\section{Introduction\label{sec:Introduction}}

Since decades noble metal nanoclusters have been attracting scientists
by imposing new interesting questions for fundamental research as
well as by promising novel opportunities for applications in nanotechnology
\citep{landman99,Shipway2000,vbk01,peyser2001,Zheng2004,Lal2007,Johnson2009,Benson2011,Lang2012}.
The possible fields of applications of noble metal nanoclusters and
their assemblies being developed nowadays range from single-molecule
probing \citep{Seydack2005,Anker2008} and medical diagnostics \citep{Haes2004,Hahn2011,Sailor2012}
to nonlinear light sources \citep{Zheng2004,Li2009,Polina2011,Polyushkin2014}
and energy transport \citep{Quinten1998,Maier2001,Maier2003,Palacios2014}.
A great deal of attention of the researchers working in this field
is paid to noble metal nanoparticles with sizes ranging form tens
to several hundreds of nanometers. These particles exhibit strong
absorption of light mainly in the visible region due to collective
excitation of conduction electrons, known as plasmons \citep{elsayed_1999,Jensen2000},
which is absent in the corresponding bulk materials. The energy of
plasmon excitation strongly depends on shape, size and dielectric
surrounding of nanoparticles or their aggregates \citep{Kelly2003,Jain2006},
which opens up a fascinating opportunity to tune the light absorption
by adjusting these parameters. Along with that, the strong near-field
of the nanoparticle can enhance the optical response of a molecule
at its vicinity \citep{Nie1997,Tam2007} or excite neighboring nanoparticles,
leading to energy transport \citep{Maier2001}.

The challenging problem of delivering excitation to a desired spatial
point at a desired instant of time using noble metal nanoparticles
and their aggregates has been recently addressed both experimentally
and theoretically, providing impressive and promising results. Using
phase modulation of the exciting laser pulse, the possibility of coherent
control of ultrafast energy localization in nanostructures has been
theoretically demonstrated \citep{Stockman2002}. Optical near field
manipulation on a sub-diffraction length scale and on a sub-picosecond
time scale has been experimentally achieved in specifically designed
aggregates constructed of silver nanoparticles via excitation by polarization-shaped
laser pulses designed using adaptive control techniques \citep{Brixner2005,Aeschlimann2010}.
Along with localization of excitation, possibilities of control of
light propagation in nanoparticle arrays has been demonstrated both
by theory \citep{Sukharev2006,Sukharev2007,Tuchscherer2009} and experiment
\citep{Aeschlimann2007}.

Aiming to further reduce the size of possible nanooptical devices
one ultimately reaches the size regime where each atom of a nanoparticle
counts. In this size range the plasmonic absorption band transforms
into molecular-like discrete energy levels which strongly depend on
the number of atoms in the cluster and their geometric structure \citep{Zheng2004,Morton2011}.
Such small particles possess properties which do not scale with the
cluster's size and are considerably different from conventional nanoparticles
as well as from corresponding bulk materials \citep{vbk01,mitric01,vbk02,Ma2012,Polina2012}.
This makes the molecular-like noble metal clusters promising building
blocks for ultra-small optical devices.

Energy localization and light propagation processes in aggregates
of the sub-nanometer noble metal clusters are, at the moment, not
studied so intensively as in case of their bigger counterparts. The
theoretical methods such as the finite-difference time-domain method
(FDTD) \citep{Maier2003,Sukharev2006}, the extended Mie theory \citep{Muehlig2011,Rockstuhl2011,Willingham2011,Solis2012,Muehlig2013},
the discrete-dipole approximation (DDA) \citep{Wang2009}, the boundary
elements method (BEM) \citep{Brixner2005}, and the quasi-static approximation
to Maxwell's equations \citep{Stockman2004} which are successful
in describing 10-100 nm sized nanoparticles, cannot be straightforwardly
applied to sub-nanometer clusters due to the small sizes and the intrinsic
quantum nature of the latter. Coupling of these methods to the Bloch
\citep{Ziolkowski1995,Fratalocchi2008}, Schrödinger \citep{Lopata2009},
and Liouville \citep{Sukharev2011} equations allowed for inclusion
of quantum effects and description of atomic systems interacting with
light. Within these approaches atoms are usually treated as two-,
three-, or four-level quantum systems with degenerate excited states.
Unfortunately, these methods are not sufficient to describe the electronic
structure of a noble metal nanocluster realistically and to simulate
the interaction of cluster aggregates with light, which is essential
for interpretation of experimental results and design of novel nanosystems.

Recently, we have proposed a method to describe light propagation
in arrays consisting of small noble metal clusters \citep{Polina2014}.
The method is based on numerical integration of the time-dependent
Schrödinger equations (TDSE) in the manifold of electronic eigenstates
of each separate cluster under the action of an external laser field
as well as the electric field produced by the other clusters in the
array. The TDSE is solved in a self-consistent iterative manner until
the convergence of time-dependent dipole moments of all clusters is
reached. The results provided by this method have been tested against
quantum-mechanical calculations and good agreement has been demonstrated
provided the distance between the clusters in the array is large enough.

In the current contribution, we combine the previously introduced
theoretical approach \citep{Polina2014} with the optimal control
employing a genetic algorithm to design phase-shaped laser pulses
which drive selectively the excitation to a selected branch of a model
T-shaped structure constructed of atomic silver clusters. We simulate
the spatio-temporal distribution of the electric field produced by
the structure under the action of optimal laser pulses. Analysis of
single cluster electron population dynamics allows for describing
the mechanism which governs the excitation transfer in the selected
direction.

The paper is structured as follows: In Sec. \ref{sec:Theory} the
theoretical approach is briefly outlined. The results are presented
and discussed in Sec. \ref{sec:Results-and-discussion}, and in Sec.
\ref{sec:Conclusion} the conclusions and outlook are given.

\section{Theory\label{sec:Theory}}

\subsection{Simulation of light propagation}

We consider here arrays consisting of $N$ metal clusters with their
charge centers placed at positions $\mathbf{R}_{I}$. We assume that
the distance between the clusters is large enough to use the dipole-dipole
approximation for modeling the cluster-cluster interaction and to
neglect the possibility of charge transfer between the clusters. These
approximations have been critically examined and justified in our
previous work \citep{Polina2014} (additionally, see the Supplemental
Material \citep{SM}). Within this approach, the Hamiltonian operator of the $I$-th
cluster can be written in the following way:

\begin{equation}
\mathrm{H}_{I}=\mathrm{H}_{I}^{0}-\boldsymbol{\mu}_{I}\cdot\left(\underset{{\scriptscriptstyle J\neq I}}{\sum}\boldsymbol{\boldsymbol{\varepsilon}}_{J}\left(\mathbf{r},t\right)+\boldsymbol{\boldsymbol{\varepsilon}}_{ext}\left(\mathbf{r},t\right)\right).\label{eq:single-cluster Hamiltonian}
\end{equation}
Here, $\mathrm{H}_{I}^{0}$ is the field-free electronic Hamiltonian
of the $I$-th cluster, $\boldsymbol{\mu}_{I}$ is the electronic
dipole moment operator of the cluster, $\boldsymbol{\boldsymbol{\varepsilon}}_{ext}\left(\mathbf{r},t\right)$
is the external electric field strength and $\boldsymbol{\boldsymbol{\varepsilon}}_{J}\left(\mathbf{r},t\right)$
is the electric field produced by the electromagnetic response of
the $J$-th cluster in the array. In this Hamiltonian, the term $-\boldsymbol{\mu}_{I}\cdot\underset{{\scriptscriptstyle J\neq I}}{\sum}\boldsymbol{\boldsymbol{\varepsilon}}_{J}\left(\mathbf{r},t\right)$
represents the cluster-cluster interaction, which in our approach
is assumed to be purely electromagnetic. The term $-\boldsymbol{\mu}_{I}\cdot\boldsymbol{\boldsymbol{\varepsilon}}_{ext}\left(\mathbf{r},t\right)$
describes the interaction with an external laser field.

In the described approach, we consider metal clusters of arbitrary
shape and size, with the only restriction that the individual components
are much smaller than the wavelength of the light used for excitation,
typically in a low nanometer size range. This allows us to consider
the external field to be uniform over the extent of a single cluster
and to represent each individual component as a dipole emitter. Thus
the terms in the Hamiltonian \eqref{eq:single-cluster Hamiltonian}
can be further reduced to $-\boldsymbol{\mu}_{I}\cdot\boldsymbol{\boldsymbol{\varepsilon}}_{ext}\left(\mathbf{R}_{I},t\right)$
for the interaction with the external field and $-\boldsymbol{\mu}_{I}\cdot\boldsymbol{\boldsymbol{\varepsilon}}_{J}(\mathbf{R}_{I},t)$
for the cluster-cluster interaction.

Employing these ingredients, the Hamiltonian of the whole cluster
array can be constructed, and the electromagnetic response can be
accurately simulated \citep{Polina2014}. The major obstacle is the
size of the Hamiltonian matrix of the whole array, which grows exponentially
with the number of clusters, thus making the calculations cumbersome
if a sufficient number of electronic states needs to be included.
To overcome this difficulty, an iterative approach to describe electron
dynamics in such systems has been developed, which allows one to perform
simulations for relatively large arrays with many electronic states
per cluster included. The applicability of this approach to the systems
constructed of Ag$_{8}$ clusters placed at relatively large distances,
as we deal with in the current work, has been demonstrated previously
\citep{Polina2014}. Here, we briefly outline the major steps of this
method.

Instead of dealing with the time-dependent wave function describing
the time evolution of the whole array, time-dependent wave functions
for each single cluster $\left|\Phi_{I}(t)\right\rangle $ are determined
separately by solving numerically the TDSE's with single-cluster Hamiltonian
$\mathrm{H}_{I}$ \eqref{eq:single-cluster Hamiltonian}. The single-cluster
wave function is expanded in the basis spanned by the eigenfunctions
$\left|\Psi_{i}^{(I)}\right\rangle $ of the field-free Hamiltonian
$\mathrm{H}_{I}^{0}$ 
\begin{equation}
\left|\Phi_{I}(t)\right\rangle =\underset{{\scriptstyle i}}{\sum}C_{i}^{(I)}(t)e^{-iE_{i}^{(I)}t}\left|\Psi_{i}^{(I)}\right\rangle ,\label{eq:wavefun iter}
\end{equation}
where $E_{i}^{(I)}$ is the $i$-th electronic state energy of the
$I$-th cluster and $C_{i}^{(I)}(t)$ is the time-dependent expansion
coefficient. The final set of coupled differential equations for the
time-dependent expansion coefficients $C_{i}^{(I)}(t)$ for numerical
integration reads:

\[
\dot{C}_{i}^{(I)}(t)=i\boldsymbol{\boldsymbol{\varepsilon}}_{ext}\left(\mathbf{R}_{I},t\right)\cdot\underset{{\scriptstyle j}}{\sum}C_{j}^{(I)}(t)e^{-i\left(E_{j}^{(I)}-E_{i}^{(I)}\right)t}\boldsymbol{\mu}_{ij}^{(I)}+
\]

\begin{equation}
i\underset{{\scriptstyle j}}{\sum}C_{j}^{(I)}(t)e^{-i\left(E_{j}^{(I)}-E_{i}^{(I)}\right)t}\boldsymbol{\mu}_{ij}^{(I)}\cdot\underset{{\scriptstyle J\neq I}}{\sum}\boldsymbol{\boldsymbol{\varepsilon}}_{J}(\mathbf{R}_{I},t).\label{eq:TDSE iter coeff}
\end{equation}

The essential quantities needed for solving the set of equations \ref{eq:TDSE iter coeff}
are the electronic state energies $E_{i}^{(I)}$ and the transition
dipole moments $\boldsymbol{\mu}_{ij}^{(I)}$ between all electronic
states included in the simulations. In principle, for molecular-sized
clusters, these quantities can be obtained using any \emph{ab initio}
or semiempirical electronic structure method. In the current work
we have used the linear response time-dependent density functional
theory (TDDFT) due to its efficiency and applicability to relatively
large complex systems. While the electronic state energies and transition
dipole moments between the ground and excited electronic states can
be obtained employing standard TDDFT routines, the construction of
full dipole coupling matrix requires an additional approximate procedure
presented in detail elsewhere \citep{werner10}.

Notably, Eq. \eqref{eq:TDSE iter coeff} contains the electric field
$\boldsymbol{\boldsymbol{\varepsilon}}_{J}(\mathbf{R}_{I},t)$ produced
by the $J$-th cluster, which can be calculated on the basis of the
time-dependent dipole moment of that cluster \citep{feynman1963},
while the latter can be determined as the expectation value of the
respective dipole moment operator.

Thus Eqs. \ref{eq:TDSE iter coeff} are coupled not only by the explicit
presence of the expansion coefficients $C_{i}^{(J)}(t)$, but also
by the implicit dependence of the electric field $\boldsymbol{\boldsymbol{\varepsilon}}_{J}(\mathbf{R}_{I},t)$
on these coefficients. Therefore the system of equations \eqref{eq:TDSE iter coeff}
has to be solved for all subunits simultaneously in the self-consistent
manner.

Consequently, our approach involves the following steps:

(i) In the first step, we determine the initial guess for the expansion
coefficients $\left\{ C_{i}^{(I)}(t)\right\} _{0}$ by solving the
set of Eqs. \eqref{eq:TDSE iter coeff} taking into account interaction
with the external electric field only $\left(\boldsymbol{\boldsymbol{\varepsilon}}_{J}(\mathbf{R}_{I},t)=0\right)$.

(ii) The first approximation to the time-dependent dipole moments
$\left\{ \mathbf{p}_{J}(t)\right\} _{0}$ of all clusters in the array
is obtained.

(iii) The response of all clusters is determined by calculating $\left\{ \boldsymbol{\boldsymbol{\varepsilon}}_{J}(\mathbf{R}_{I},t)\right\} _{0}$
and used in the set of Eqs. \eqref{eq:TDSE iter coeff} to find the
next approximation to the expansion coefficients $\left\{ C_{k}^{(I)}(t)\right\} _{1}$.

Steps (ii)-(iii) are repeated until convergence is reached. Since
in the simulations the essential quantities are the time-dependent
dipole moments, the criterion for convergence is that the difference
between the dipole moments obtained in subsequent iterations $i$
and $i+1$ integrated over the whole simulation time $T$ is less
than a certain threshold: 
\begin{equation}
\delta=\frac{1}{T}\underset{{\scriptstyle {\scriptscriptstyle J}}}{\sum}\underset{{\scriptstyle {\scriptscriptstyle 0}}}{\overset{{\scriptstyle {\scriptscriptstyle T}}}{\int}}\left|\mathbf{p}_{J}^{i+1}(t)-\mathbf{p}_{J}^{i}(t)\right|\mathrm{d}t<\epsilon.\label{eq:deviation}
\end{equation}
The convergence of the method is usually reached due to the fact that
the main contribution to the coupling between the electronic states
of a single cluster comes from the external laser field, and the interaction
with the electric field produced by other clusters brings only small
perturbation. Thus the initial guess obtained in the step (i) is already
a good approximation to the final results, which is further corrected
by the inclusion of cluster-cluster interactions.

When the calculation converges, the single-cluster time-dependent
dipole moments are determined using the expansion coefficients found
and the spatial-temporal electric field distribution is calculated
according to \citep{feynman1963}

\[
\mathbf{E}\left(\mathbf{r},t\right)=\underset{{\scriptscriptstyle I}}{\overset{{\scriptscriptstyle N}}{\sum}}\left\{ \frac{-1}{r_{I}^{3}}\left[\mathbf{p}_{I}\left(t'\right)+\frac{r_{I}}{c}\mathbf{\dot{p}}_{I}\left(t'\right)-\frac{3\mathbf{r}_{I}}{r_{I}^{2}}\left(\mathbf{r}_{I}\cdot\left(\mathbf{p}_{I}\left(t'\right)+\frac{r_{I}}{c}\mathbf{\dot{p}}_{I}\left(t'\right)\right)\right)+\right.\right.
\]

\begin{equation}
\left.\left.\frac{1}{c^{2}}\mathbf{r}_{I}\times\left(\mathbf{\ddot{p}}_{I}\left(t'\right)\times\mathbf{r}_{I}\right)\right]_{t'=t-\frac{r_{I}}{c}}\right\} ,\label{eq:field sing}
\end{equation}
where the summation runs over all clusters in the array, $\mathbf{r}_{I}=\mathbf{r}-\mathbf{R}_{I}$
and $r_{I}$ is its absolute value.

\subsection{Optimal control of light propagation\label{sub:Optimal-control}}

\begin{figure}[h]
\begin{centering}
\includegraphics[width=0.5\columnwidth]{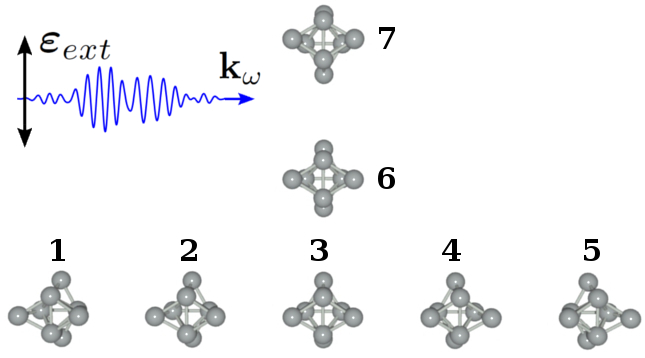} 
\par\end{centering}

\protect\caption{A T-shaped structure consisting of seven Ag$_{8}$ clusters placed
at 20 a$_{0}$ separation irradiated by an external laser pulse $\boldsymbol{\boldsymbol{\varepsilon}}_{ext}\left(\mathbf{r},t\right)$
with the wave vector $\mathbf{k}_{\omega}$.}

\label{Fig:array scheme} 
\end{figure}

The aim of the control simulations is to localize the electric field
in a specified time interval around a particular spatial part of the
nanostructure. In the present contribution, we wish to control the
light propagation and localization in a T-shaped metal cluster array
consisting of seven identical Ag$_{8}$ clusters placed at 20 a$_{0}$
distance between closest neighbors and irradiated by a phase-shaped
short laser pulse (see Fig. \ref{Fig:array scheme}).

Thus, as a measure of the spatio-temporal electric field localization
we define the following function for each member of the cluster array:

\begin{equation}
W_{I}=\frac{1}{\tau}\underset{{\scriptstyle {\scriptscriptstyle T}-\tau}}{\overset{{\scriptstyle {\scriptscriptstyle T}}}{\int}}\underset{{\scriptstyle {\scriptscriptstyle 0}}}{\overset{{\scriptstyle {\scriptscriptstyle R_{s}}}}{\int}}\underset{{\scriptscriptstyle 4\pi}}{\iint}\left|\mathbf{E}\left(\mathbf{r}'+\mathbf{R}_{I},t\right)\right|^{2}\mathrm{d}r'\mathrm{d}\Omega\mathrm{d}t,\label{eq:field energy}
\end{equation}
where $T$ is the simulation time and $\tau$ is the temporal localization
interval. The spatial integration is performed within a sphere with
the radius $R_{s}$ centered at the specified cluster. In the optimal
control simulations we minimize the ratio of the $W$-functions for
specified cluster pairs.

A time-dependent phase-shaped external laser pulse is introduced in
our model as:

\begin{equation}
\boldsymbol{\boldsymbol{\varepsilon}}_{ext}\left(\mathbf{r},t\right)=\frac{1}{2}\underset{\omega}{\sum}S\left(\omega\right)\left(e^{i\left(\omega t-\mathbf{k}_{\omega}\cdot\mathbf{r}+\varphi\left(\omega\right)\right)}+c.c.\right)\sin^{2}\left(\frac{\pi t}{T_{p}}\right),\label{eq:laser pulse}
\end{equation}
where $S\left(\omega\right)$ is the spectrum of the pulse, modeled
by a Gaussian function with the mean frequency $\omega_{c}$ and standard
deviation $\sigma_{\omega}$ 
\begin{equation}
S\left(\omega\right)=\frac{1}{\sqrt{2\pi}\sigma_{\omega}}e^{-\frac{\left(\omega-\omega_{c}\right)^{2}}{2\sigma_{\omega}^{2}}},\label{eq:spectrum pulse}
\end{equation}
$\mathbf{k}_{\omega}$ is the wave vector, $\varphi\left(\omega\right)$
is the spectral phase, and the sin$^{2}$-mask ensures that the pulse
fits into the pulse duration period $T_{p}\leq T$. The laser pulse
propagates along the long side of the array and is polarized in the
plane of the array, as it is shown in Fig. \ref{Fig:array scheme}.
The dependence of the spectral phase on the frequency $\omega$ is
determined as follows:

\begin{equation}
\varphi\left(\omega\right)=A\sin\left(B\left(\omega-\omega_{c}\right)+C\left(\omega-\omega_{c}\right)^{2}\right).\label{eq:spectral phase}
\end{equation}

Varying the coefficients $A$, $B$, and $C$ one can obtain different
temporal profiles of the external laser pulse. In our simulations,
we use a genetic algorithm \citep{goldberg89} to find the proper
values $A_{opt}$, $B_{opt}$, and $C_{opt}$ which generate such
external pulses that drive the excitation ``up'' (meaning $W_{7}>W_{5}$)
or ``straight'' ($W_{5}>W_{7}$) after the pulse ceases. Thus the
target functions we seek to minimize by means of the genetic algorithm
are $f_{1}=W_{5}/W_{7}$ and $f_{2}=W_{7}/W_{5}$. We used population
size of 30 ``species'' and ran the optimization for 30 generations
to determine the optimal set of parameters $A_{opt}^{1}$, $B_{opt}^{1}$,
and $C_{opt}^{1}$ which minimize $f_{1}$. Afterwards, in the same
manner the set of parameters $A_{opt}^{2}$, $B_{opt}^{2}$, and $C_{opt}^{2}$
minimizing $f_{2}$ has been determined.

\section{Results and discussion\label{sec:Results-and-discussion}}

First the equilibrium structure of Ag$_{8}$ cluster has been determined
by the full geometry optimization in the ground electronic state employing
DFT with the gradient corrected Becke-Lee-Yang-Parr (BLYP) exchange-correlation
functional \citep{becke88,lyp}, combined with the triple zeta valence
plus polarization Gaussian atomic basis set (TZVP) \citep{tzvp} and
a relativistic 19-electron effective core potential for silver \citep{andrae1990}.
Subsequently, a number of excited electronic state energies $E_{i}^{(I)}$
has been calculated employing linear response TDDFT (LR-TDDFT). All
calculations have been performed using the TURBOMOLE V.6.3 package
\citep{turbomole,TurbomoleTDDFTEx}. Then, the transition dipole moments
$\boldsymbol{\mu}_{ij}^{(I)}$ between all the electronic states were
determined according to Ref. \citep{werner10}. The most intense transition
in the visible and near-UV range has an excitation energy of 3.8 eV.
Therefore the central frequency of the external laser pulse is set
to $\omega_{c}$=3.8 eV to be resonant with this intense cluster transition.
The parameters $\sigma_{\omega}$ and $T_{p}$ of the external laser
pulse as well as other computational details are given in the Supplemental
Material \citep{SM}.

\begin{figure}[h]
\begin{centering}
\includegraphics[width=0.5\columnwidth]{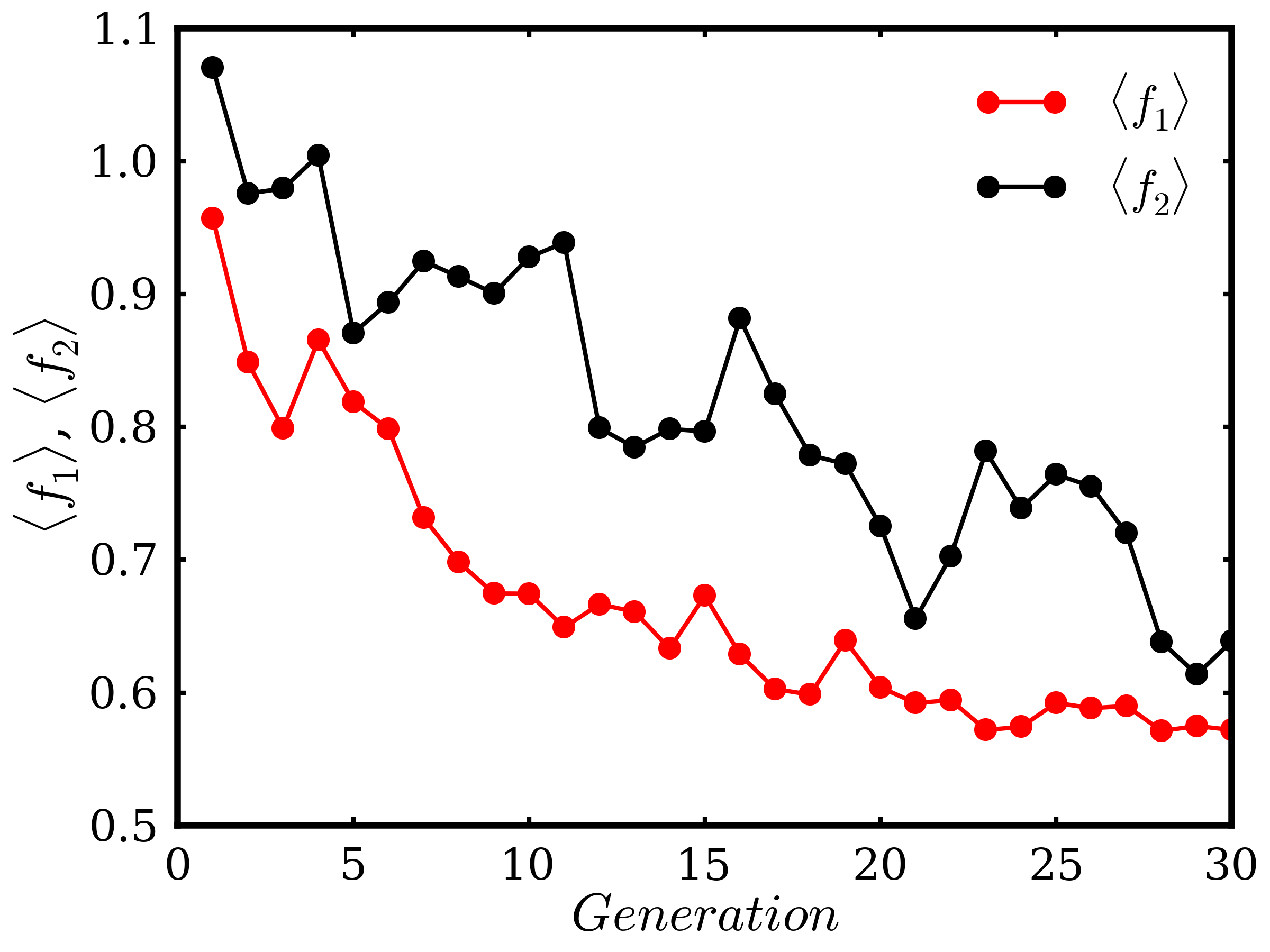} 
\par\end{centering}

\protect\caption{Values of the target functions $f_{1}$ and $f_{2}$ at each generation,
averaged over all ``species'' within this generation.}

\label{Fig:f1-f2 optimization} 
\end{figure}

We have performed optimal control simulations with the goal to steer
the electric field in two different directions along the T-shaped
cluster array shown in Fig. \ref{Fig:array scheme}. In Fig. \ref{Fig:f1-f2 optimization}
we present the dependence of the target function on the generation
number showing the convergence of the optimization algorithm. Each
point corresponds to the values of the target functions $f_{1}$ and
$f_{2}$ defined in Sec. \ref{sub:Optimal-control} and averaged over
all ``species'' within this generation. The minimal value of $\left\langle f_{1}\right\rangle $
of 0.57 was obtained in the last generation, while the minimal value
of $\left\langle f_{2}\right\rangle $ being equal to 0.63 was obtained
in the one-to-last generation. The optimal laser pulse P1 was reconstructed
in time domain using Eqs. \eqref{eq:laser pulse}-\eqref{eq:spectral phase}
with the set of parameters giving rise to the minimal value of $f_{1}$
over all ``species'' ($A_{opt}^{1}$=6.15, $B_{opt}^{1}$=-16.19
eV$^{-1}$, and $C_{opt}^{1}$=-101.79 eV$^{-2}$), and for the optimal
laser pulse P2 the set of parameters corresponding to the minimal
value of $f_{2}$ over all ``species'' is used ($A_{opt}^{2}$=5.95,
$B_{opt}^{2}$=-16.14 eV$^{-1}$, and $C_{opt}^{2}$=-18.35 eV$^{-2}$).

\begin{figure}[t]
\begin{centering}
\includegraphics[width=1\textwidth]{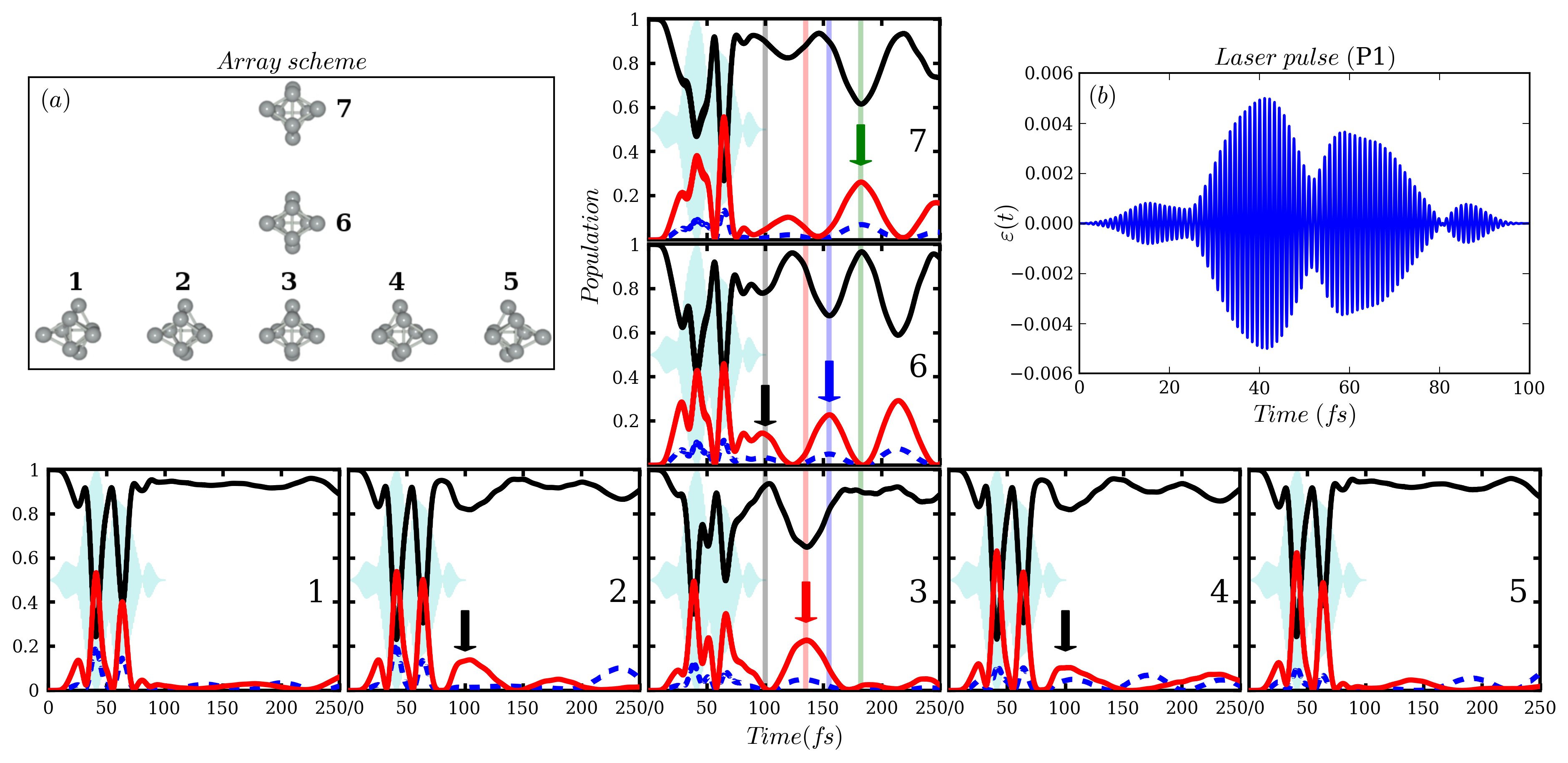} 
\par\end{centering}

\protect\caption{Electron population dynamics of single clusters in the array under
the action of the laser pulse P1. Population of the ground state $\left|C_{0}^{(I)}\right|^{2}$
is shown with black lines, of the S$_{17}$ excited state $\left|C_{17}^{(I)}\right|^{2}$
with blue lines, and of the S$_{18}$ excited state $\left|C_{18}^{(I)}\right|^{2}$
with red lines. The subplots are arranged and numbered according to
the array scheme which is presented in the inset (a). The temporal
profile of the external laser pulse P1 is shown in the inset (b).
The vertical arrows denote different steps of the excitation propagation
``up'' along the side chain of the array. }

\label{Fig:p1 dynamics} 
\end{figure}

The electronic state population dynamics of each cluster in the array
under the action of the external laser pulse P1 is presented in Fig.
\ref{Fig:p1 dynamics}. The subplots with population dynamics are
arranged in order to reproduce the location of the cluster in the
array and numbered according to the array scheme shown in the inset
(a) for convenience. The temporal profile $\boldsymbol{\boldsymbol{\varepsilon}}\left(t\right)$
of the laser pulse P1 is presented in the inset (b). The ground state
population $\left|C_{0}^{(I)}\right|^{2}$ of each single cluster
is plotted with the black lines and the populations of the dominantly
excited states $\left|C_{17}^{(I)}\right|^{2}$ and $\left|C_{18}^{(I)}\right|^{2}$
is shown with blue and red lines, respectively. Since the latter two
states are degenerate and both have nonzero projections of transition
dipole moments on the laser pulse polarization vector, both states
are populated during the pulse action. It is seen, that at the end
of the pulse duration time mainly the ground state of all the clusters
is populated, but the last small peak of the laser pulse at $\sim$90
fs (cf. Fig. \ref{Fig:p1 dynamics} (b)) causes population transfer
to the excited states of the clusters 2, 4, and 6. This increase of
population is annotated on Fig. \ref{Fig:p1 dynamics} (2,4,6) with
black arrows. After the external pulse ceases the electromagnetic
interaction between clusters starts to play a decisive role in the
population dynamics. The electric field produced by the clusters 2,
4, and 6 is the strongest at the position of the cluster 3, causing
the excitation transfer to that cluster. The increase in the excited
state population of the cluster 3 is denoted in Fig. \ref{Fig:p1 dynamics}
(3) with the red arrow. Due to the configuration of the array and
the chosen orientation of the clusters, the dipole-dipole interaction
between the cluster pair 3-6 is approximately 5 times higher than
that between cluster pairs 2-3 and 3-4 in the excited state S$_{18}$.
Therefore, the excitation is propagated ``up'' along the side chain
of the array involving clusters 6 and 7. These steps are denoted in
Fig. \ref{Fig:p1 dynamics} (6) and (7) with blue and green arrows,
respectively. After the excitation reaches the last cluster in the
side chain, it is ``reflected'' back since in our simulations the
array is considered to be a closed system.

\begin{figure}[h]
\begin{centering}
\includegraphics[width=0.5\columnwidth]{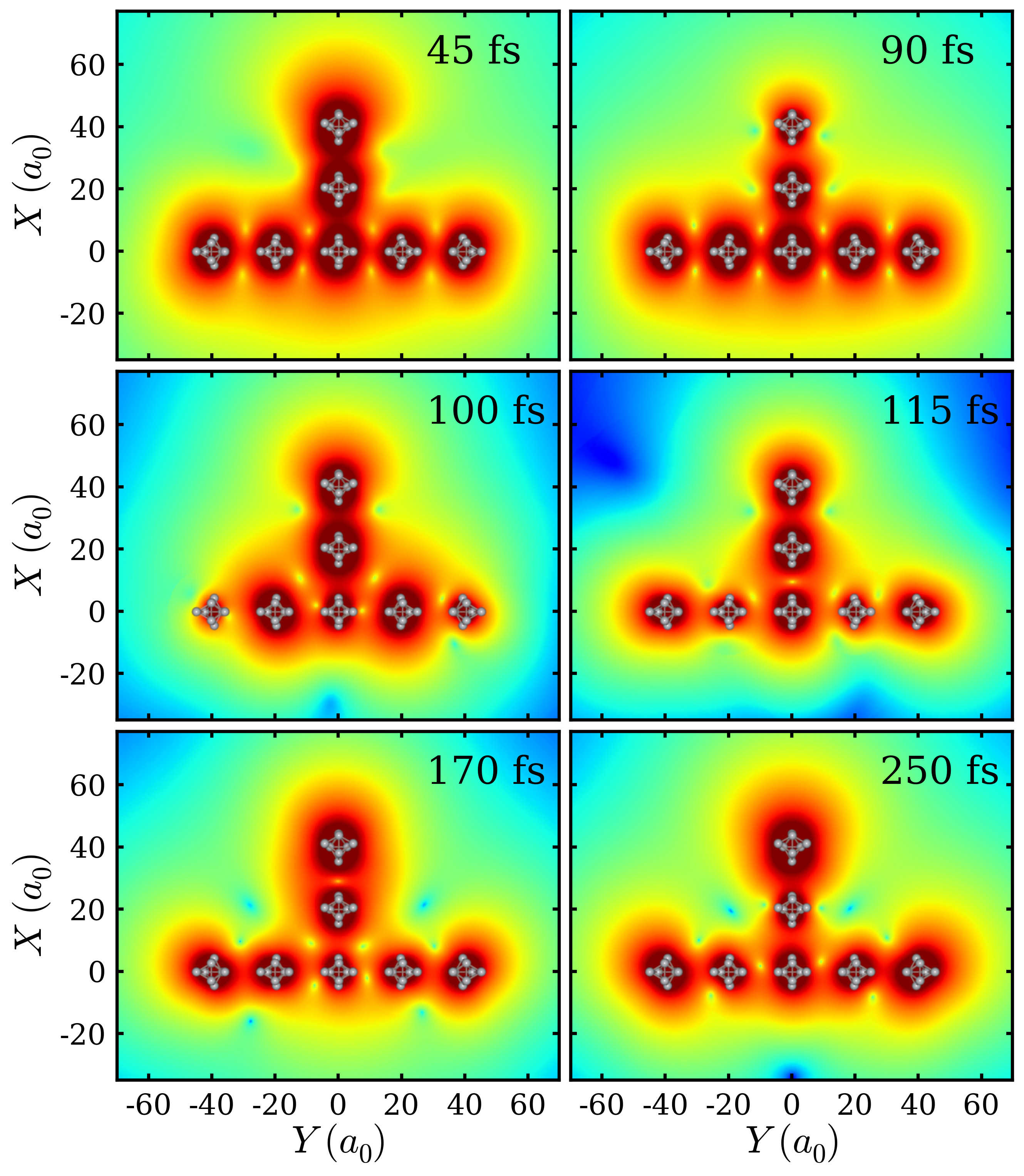} 
\par\end{centering}

\protect\caption{Spatial distribution of the electric field ($\left|\mathbf{E}\left(\mathbf{r},t\right)\right|^{2}$)
at selected instants of the simulation time. The snapshots are taken
in the plane of the array (\textit{XY}-plane) at the time moments
which are specified in the subplots. The magnitude of the electric
field $\left|\mathbf{E}\left(\mathbf{r},t\right)\right|^{2}$ is denoted
with the color code from blue (the lowest) to red (the highest).}

\label{Fig:p1 snapshots} 
\end{figure}

The process of the excitation transfer can be better visualized by
simulations of the electric field distribution in the area around
the silver cluster array. The electric field strength is determined
employing Eq. \eqref{eq:field sing}, and the electric field energy
density is proportional to the $\left|\mathbf{E}\left(\mathbf{r},t\right)\right|^{2}$.
In Fig. \ref{Fig:p1 snapshots} the distribution of the electric field
$\left|\mathbf{E}\left(\mathbf{r},t\right)\right|^{2}$ at selected
instants of time is presented. During the external laser pulse action
the electric field varies strongly with time but does not differ much
from cluster to cluster, as it is seen for instance at 45 fs of simulation
time. When the laser pulse is about to vanish, the electric field
around the central clusters of the array is noticeably stronger than
around the side clusters. At 100 and 115 fs of the simulation time
it is clearly seen how the excitation is transferred from the clusters
2, 4 and 6 to the cluster 3, and afterwards propagates ``up'' to
clusters 6 and 7. At later instants of time the electric field is
localized mainly around the top of the array (cluster 7).

\begin{figure}[h]
\begin{centering}
\includegraphics[width=1\textwidth]{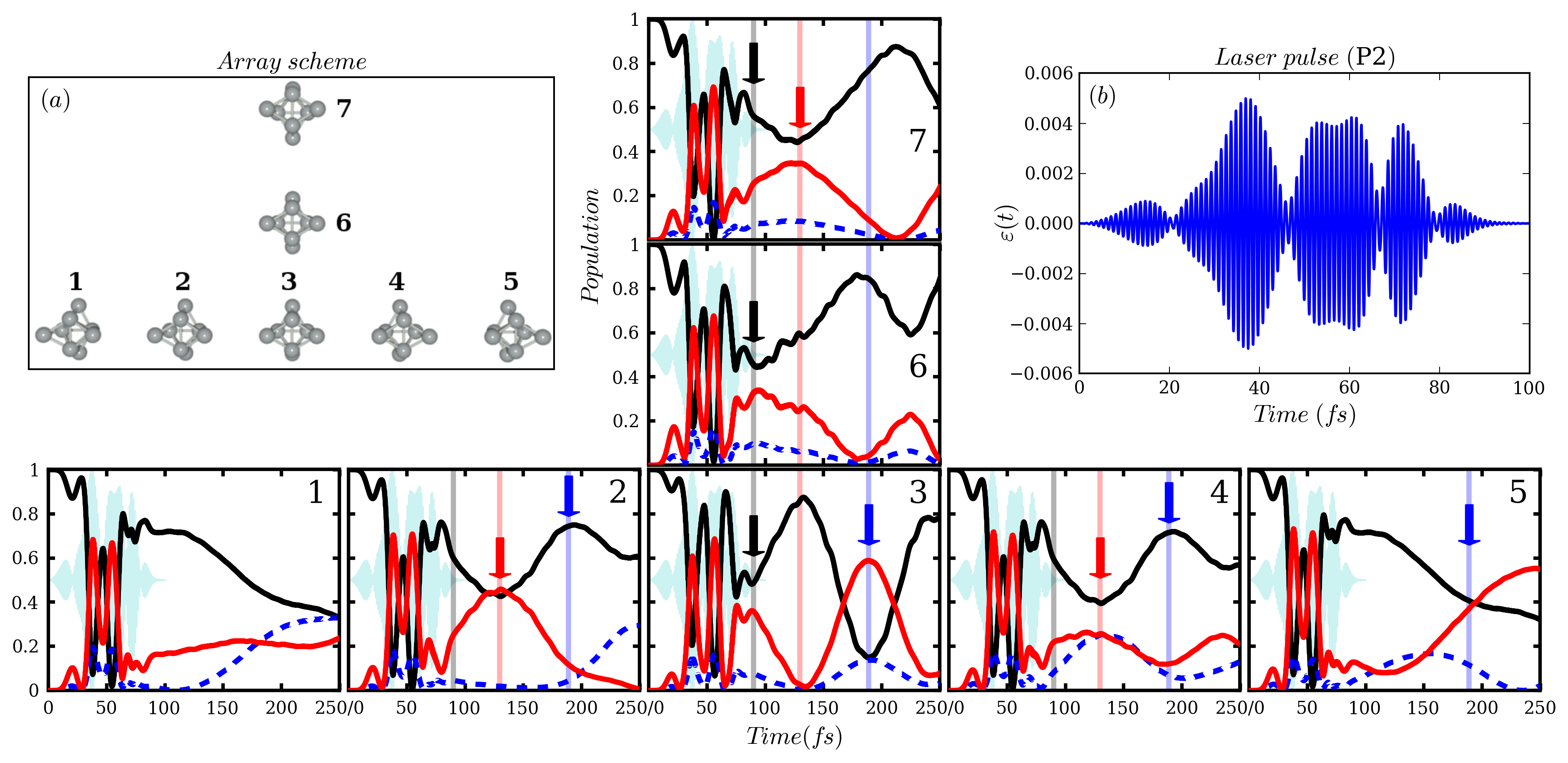} 
\par\end{centering}

\protect\caption{Electron population dynamics of single clusters in the array under
the action of the laser pulse P2. Population of the ground state $\left|C_{0}^{(I)}\right|^{2}$
is shown with black lines, of the S$_{17}$ excited state $\left|C_{17}^{(I)}\right|^{2}$
with blue lines, and of the S$_{18}$ excited state $\left|C_{18}^{(I)}\right|^{2}$
with red lines. The subplots are arranged and numbered according to
the array scheme which is presented in the inset (a). The temporal
profile of the external laser pulse P2 is shown in the inset (b).
The vertical arrows denote different steps of the excitation propagation
``straight'' along the horizontal side of the array. }

\label{Fig:p2 dynamics} 
\end{figure}

In the following, we discuss the electron population dynamics of the
silver cluster array and the spatio-temporal distribution of the electric
field under the action of the second external laser pulse P2, which
was optimized to drive the excitation straight along the long side
of the array. The electronic state population dynamics of each cluster
is presented in Fig. \ref{Fig:p2 dynamics}. This Figure is arranged
in the same way as Fig. \ref{Fig:p1 dynamics}. The subplots presenting
the electron population of the single clusters are placed to resemble
the spatial configuration of the silver clusters in the array and
the two insets show the array scheme (inset (a)) and the temporal
profile of the external pulse (inset (b)). In general, the dipole
coupling between the clusters 1-2-3-4-5 forming the horizontal part
of the array in the ground and the excited state S$_{18}$ is considerably
smaller than that between clusters 3-6-7 in the side chain, but for
the coupling in the excited state S$_{17}$ the situation is the opposite.
Thus to propagate the excitation ``straight'' along the horizontal
side of the array it is required to selectively populate the excited
state S$_{17}$, which is not strongly populated by the external laser
pulse due to the small projection of the corresponding transition
dipole moment on the polarization direction of the pulse. Consequently,
in the optimization procedure, the pulse P2 has been designed, which
leads to more complicated electron dynamics as compared to the pulse
P1. The action of P2 results in a decrease of the ground state population
of the central clusters of the array (2,3,4 and 6) down to 0.5-0.6
(shown in Fig. \ref{Fig:p2 dynamics} (3), (6), and (7) with black
arrows). After the pulse ceases, due to the strong coupling between
the clusters 3 and 6 the excitation is rapidly transferred to the
cluster 6, and then some part is further propagated ``up'' to the
cluster 7. However, due to the high electron population in the excited
state S$_{18}$ of the cluster 6, the interaction between clusters
2-6 and 4-6 starts to play a role for transferring the population
to the excited states S$_{18}$ of the cluster 2 as well as S$_{17}$
and S$_{18}$ of the cluster 4. This step is marked in Fig. \ref{Fig:p2 dynamics}
(2), (4), and (7) with red arrows. Finally, part of the excitation
mainly from the state S$_{17}$ of the cluster 4 is propagated ``straight''
along the long side of the array to the cluster 5 leading to monotonic
increase of the electron population of the S$_{18}$ state . The rest
of the excitation is redistributed among the other clusters in this
part of the array. This step is annotated in Fig. \ref{Fig:p2 dynamics}
(2)-(5) with blue arrows.

\begin{figure}[h]
\begin{centering}
\includegraphics[width=0.5\columnwidth]{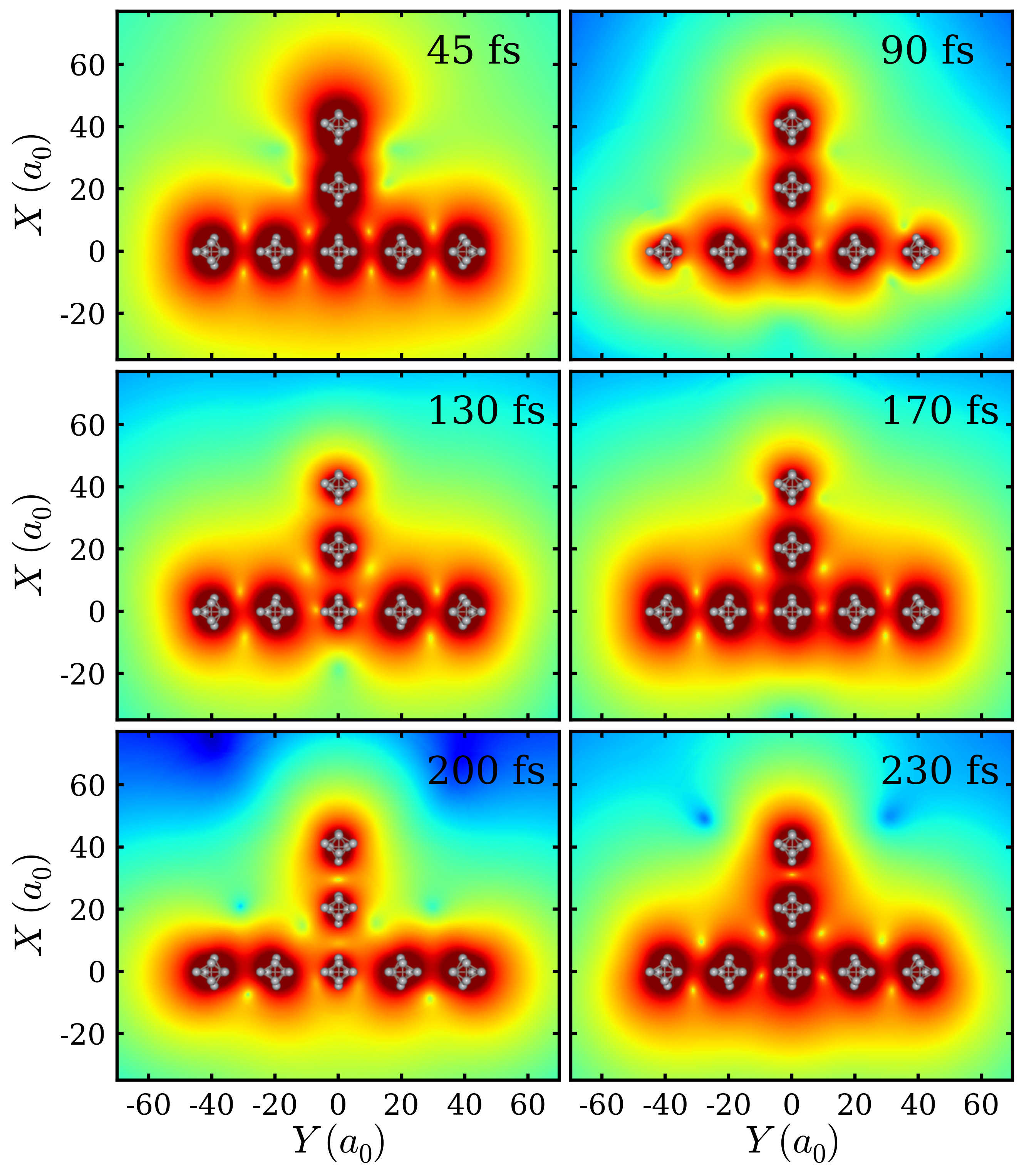} 
\par\end{centering}

\protect\caption{Spatial distribution of the electric field ($\left|\mathbf{E}\left(\mathbf{r},t\right)\right|^{2}$)
at selected instants of the simulation time. The snapshots are taken
in the plane of the array (\textit{XY}-plane) at the time moments
which are specified in the subplots. The magnitude of the electric
field $\left|\mathbf{E}\left(\mathbf{r},t\right)\right|^{2}$ is denoted
with the color code from blue (the lowest) to red (the highest).}

\label{Fig:p2 snapshots} 
\end{figure}

The spatio-temporal variation of the electric field produced by the
array is illustrated in Fig. \ref{Fig:p2 snapshots}. As in the case
of P1 excitation, during the pulse action the electric field does
not differ much for different clusters. At the end of the excitation
time ($\sim$90 fs) the electric field is noticeably stronger in the
central part of the array. After the external pulse ceases, the electric
field is redistributed around the clusters that form the long side
of the array (see Fig. \ref{Fig:p2 snapshots} 130-230 fs).

\begin{figure}[h]
\begin{centering}
\includegraphics[width=0.5\columnwidth]{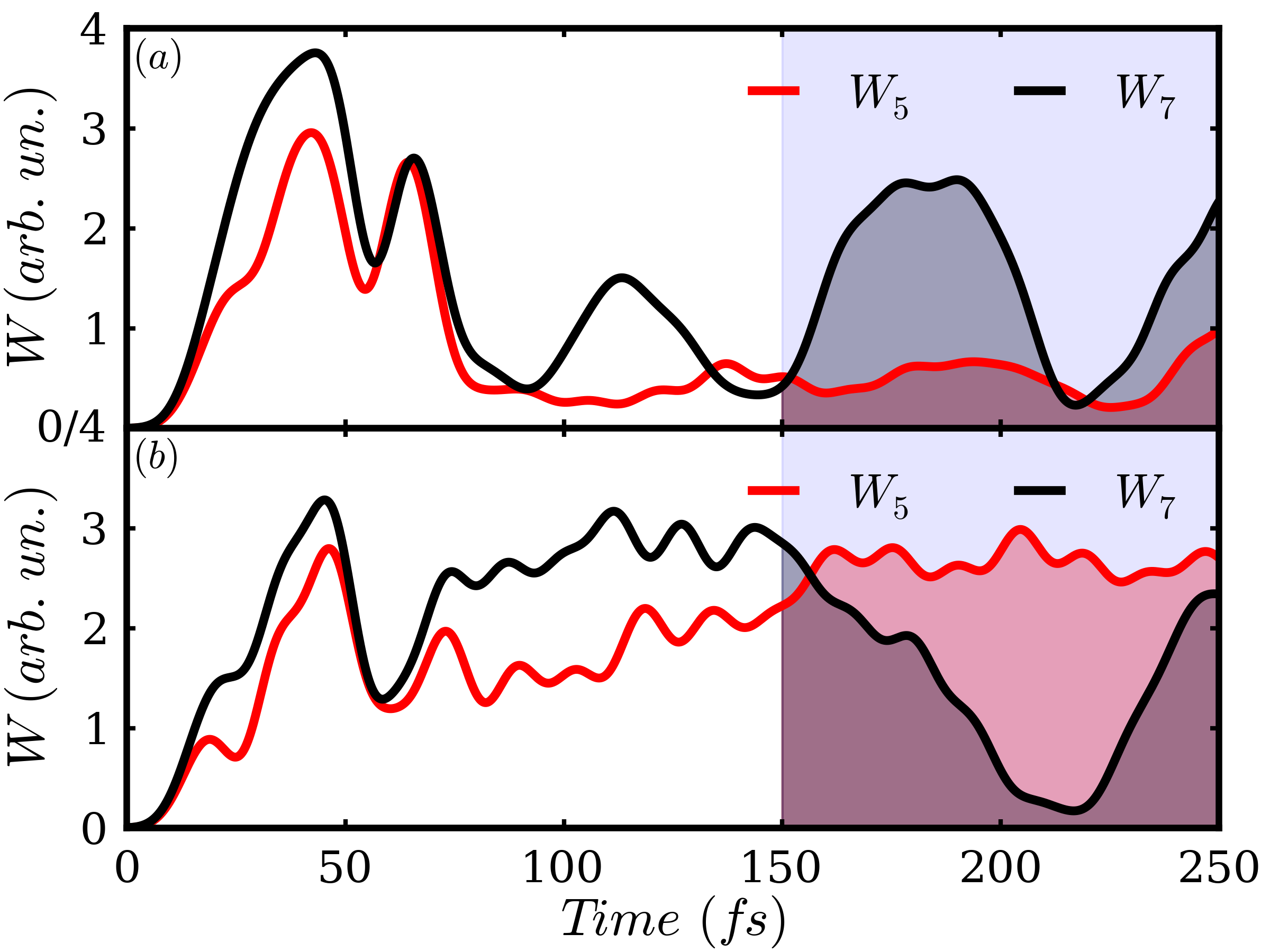} 
\par\end{centering}

\protect\caption{Electric field energies $W_{I}\left(t\right)$ calculated using Eq.
\eqref{eq:field energy} for cluster 5 ($W_{5}$) and cluster 7 ($W_{7}$).
The results are obtained for the silver cluster array excitation with
(a) the laser pulse P1, and (b) the laser pulse P2. The time interval
of optimization of the ratio between these energies is shaded in both
pictures.}

\label{Fig:p1-p2 energies} 
\end{figure}

Finally, we compare the electric field energy around the clusters
5 and 7 along the simulation under the action of pulses P1 and P2.
As a measure of the electric field energy we use the functions $W_{5}\left(t\right)$
and $W_{7}\left(t\right)$ calculated using Eq. \eqref{eq:field energy}
without integration over time. The results are presented in Fig. \ref{Fig:p1-p2 energies}
(a) for the pulse P1 and (b) for the pulse P2. The time interval $\tau$
within which the energy has been integrated to obtain the target functions
$f_{1}$ and $f_{2}$ is shaded on the figures. It is clearly seen
in Fig. \ref{Fig:p1-p2 energies} (a) that within the interval of
interest the electric field energy propagated to the ``top'' of
the array (to the cluster 7) significantly exceeds its counterpart
localized around cluster 5. In other words, most of the electric field
energy is propagated ``up'' along the side chain of the array. On
the other hand, after the action of the pulse P2 the electric field
is much stronger around cluster 5 than around cluster 7, i.e. the
electric field is propagated ``straight'' along the long side of
the array. This demonstrates that selective switching of the light
localization can be achieved by analytically parametrized optimal
laser fields, which is of interest for the development of plasmonic
nanoelectronic devices with ultrafast optical response.

\section{Conclusion\label{sec:Conclusion}}

We have demonstrated the possibility to optimally control the light
propagation in a T-shaped nanostructure built up of small atomic silver
clusters. For this purpose we have combined our recently developed
iterative method for the simulation of light propagation in metal
cluster arrays with the optimal control using a genetic algorithm.
The driving field was analytically parametrized in the frequency domain.
The ground and excited state energies of individual clusters as well
as the transition dipole moments required for the electron dynamics
simulation are determined based on \emph{ab initio} TDDFT calculations,
which allowed for a realistic description of the electronic structure
of the clusters. We have demonstrated that laser pulses can be designed
which selectively drive the electric field either along the main axis
of the array or along the side chain. This allows to achieve spatio-temporal
electric field localization in different parts of the nanostructure
over time intervals of \textasciitilde{} 100 fs. Thus our ``proof
of principle'' simulations demonstrate that nanoelectronic devices
build from small noble metal clusters can operate selectively in an
ultrafast regime and thus may serve as building blocks for the next
generation of nanoplasmonic devices.

\section*{Acknowledgments}

The authors are grateful to financial support by the DFG SPP 1391
``Ultrafast Nanooptics'' (Grant Number MI-1236). R.M. acknowledges the support in the frame of the ERC Consolidator Grant Project DYNAMO.

\providecommand*{\mcitethebibliography}{\thebibliography} \csname
@ifundefined\endcsname{endmcitethebibliography} {\let\endmcitethebibliography\endthebibliography}{}
 
\end{document}